\documentclass[%
reprint,
longbibliography,
amsmath,amssymb,
aps,
pra,
]{revtex4-1}
\usepackage{graphicx}% Include figure files
\usepackage{dcolumn}% Align table columns on decimal point
\usepackage{bm}% bold math
\usepackage{afterpage}
\begin{document}

\preprint{APS/123-QED}

\title{Hofstadter Butterfly Evolution in the Space of Two-Dimensional Bravais Lattices}% Force line breaks with \\

\author{F. Y{\i}lmaz}
\email{firat.yilmaz@bilkent.edu.tr}
\author{M. \"{O}. Oktel}
\affiliation{Department of Physics, Bilkent University, 06800, Ankara, Turkey}


%\affiliation{
%\affiliation{
%\affiliation{%


\date{\today}% It is always \today, today,

\begin{abstract}
The self-similar energy spectrum of a particle in a periodic potential under a magnetic field, known as the Hofstadter butterfly, is determined by the lattice geometry as well as the external field. Recent realizations of artificial gauge fields and adjustable optical lattices in cold atom experiments necessitate the consideration of these self-similar spectra for the most general two-dimensional lattice. In a previous work, we investigated the evolution of the spectrum for an experimentally realized lattice which was tuned by changing the unit cell structure but keeping the square Bravais lattice fixed. We now consider all possible Bravais lattices in two dimensions and investigate the structure of the Hofstadter butterfly as the lattice is deformed between lattices with different point symmetry groups. We model the optical lattice by a sinusoidal real space potential and obtain the tight binding model for any lattice geometry by calculating the Wannier functions. We introduce the magnetic field via Peierls substitution and numerically calculate the energy spectrum. The transition between the two most symmetric lattices, i.e. the triangular and the square lattice displays the importance of bipartite symmetry featuring deformation as well as closing of some of the major energy gaps. The transition from the square to rectangular and from the triangular to centered rectangular lattices are analyzed in terms of coupling of one-dimensional chains. We calculate the Chern numbers of the major gaps and Chern number transfer between bands during the transitions. We use gap Chern numbers to identify distinct topological regions in the space of Bravais lattices.
\end{abstract}
\maketitle
\renewcommand{\thefigure}{\arabic{figure}}
%\tableofcontents
\section{Introduction}
The Hofstadter butterfly is the self-similar energy spectrum of an electron moving in a periodic potential under a uniform magnetic field\cite{Hofstadter}. The self-similar behavior is due to the competition between two length scales, the lattice constant and the magnetic length. The energy spectrum and the topological properties of the system are determined solely by the magnetic field and the lattice geometry. The lattice geometry is in turn determined by the Bravais lattice and the structure of the unit cell. 

The observation of the self-similar spectrum requires a magnetic flux which is on the order of one flux quantum per unit cell. For typical solid state lattices, this requires very high magnetic fields on the order of thousands of Tesla, which is not experimentally accessible. One solution to overcome this limitation is to increase the lattice constant, such as the two stacked layers forming a superlattice with a larger lattice constant\cite{SuperlatticeKim}. A new and different approach is demonstrated in cold atom experiments utilizing artificial gauge fields\cite{EffMFJakschZoller,KetterleHarper}. Both of these approaches can create arbitrary lattices with desired lattice parameters. A striking demonstration of this flexibility is the creation of the two dimensional tunable optical lattices\cite{EsslingerDirac}. Therefore, the study of the energy spectra under a magnetic field for arbitrary lattices and their dynamical properties are relevant problems today.

In a previous work\cite{FiratNurOktel}, we calculated and examined the energy spectrum and the topological properties of the experimentally realized two-dimensional (2D) tunable optical lattices created by the Zurich group\cite{EsslingerDirac}. That lattice can be tuned from the square lattice to a Honeycomb-like geometry. However, throughout this transition, the unit cell remains square and only the unit cell potential is varied. The honeycomb-like geometry which also referred to as the brick wall lattice still has a square Bravais lattice which has a two point basis. A natural follow up question is to ask: ``How does the energy spectrum evolve as the symmetries of the underlying Bravais lattice changes?''

In this paper, we answer this question by calculating the energy spectrum for all two-dimensional Bravais lattices. We particularly investigate the transitions between lattices of distinct symmetry groups. To this end, we propose a real space sinusoidal potential which can create any arbitrary Bravais lattice with a one point basis. Such a potential can be created as an optical lattice which can be adjusted to generate all two dimensional Bravais lattices. We consider this potential in the deep lattice limit to describe the system by a tight binding (TB) model. We calculate the TB parameters by both fitting momentum space bands to numerical solution of the Schr\"{o}dinger equation and also by calculating the Wannier functions (WF).

Once the zero field TB Hamiltonian is constructed, the effect of the magnetic field is introduced by the Peierls substitution\citep{Peierls}. This method modifies the hopping amplitude with a complex phase with the constraint that the sum of the hopping phases over any closed loop on the lattice is proportional to the total magnetic flux through the loop. While this substitution only works for lattices in the TB limit, it is actually a more factual description of the recent cold atom lattice experiments where the artificial magnetic field is generated by creating these phases by modifying the tunneling between neighboring sites\citep{KetterleHarper,EffMFJakschZoller,AidelsburgerBloch}. Because the phases are written on the links between the sites, it is easier to envision changing the lattice geometry by keeping the magnetic flux per plaquette of the lattice constant.

We first reduce the magnetic TB Hamiltonian to a q-by-q matrix by using translation symmetries under a magnetic flux per plaquette which we take to be $p/q$ times the magnetic flux quantum. We restrict $p$ and $q$ to be co-prime integers. We numerically diagonalize this matrix for each $k$-point in the magnetic Brillouin zone to determine band edges. Repeating this calculation for different Bravais lattice parameters enables us to investigate the evolution of the energy spectrum.

\begin{figure*}
\centerline{\includegraphics[width=0.94\textwidth]{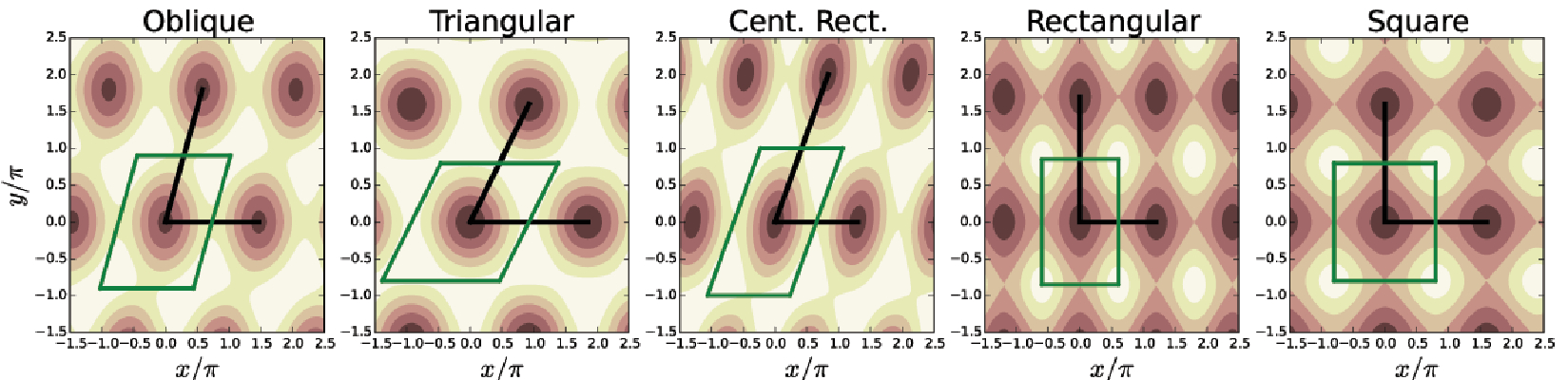}}
\caption{(Color online) Five Bravais lattices created by the potential in Eq.\ref{potentialEqn}. (from the left) Oblique, centered-rectangular(Rhombic), triangular, rectangular, square lattices. Lattice depths are $V_x = V_y = 50 (E_R),$ where $E_R$ is the recoil energy. The unit cell and the primitive vectors are shown with thick lines.}\label{2DBravaisFig}
\end{figure*}
We find that the geometry of the Bravais lattice plays a critical role in determining the energy spectrum. New symmetries can emerge according to the point group of the lattice. In two dimensions, Bravais lattices can be classified into five distinct classes by their point group. The triangular and the square lattice are the most symmetric 2D lattices. A reduction of one mirror symmetry produces the centered rectangular and rectangular lattices from them. The oblique lattice is the most general form for 2D Bravais lattices. We parametrize the space of all these lattices and observe the evolution of the Hofstadter butterfly during the transition between them.
\begin{figure}
\includegraphics[width=0.47\textwidth]{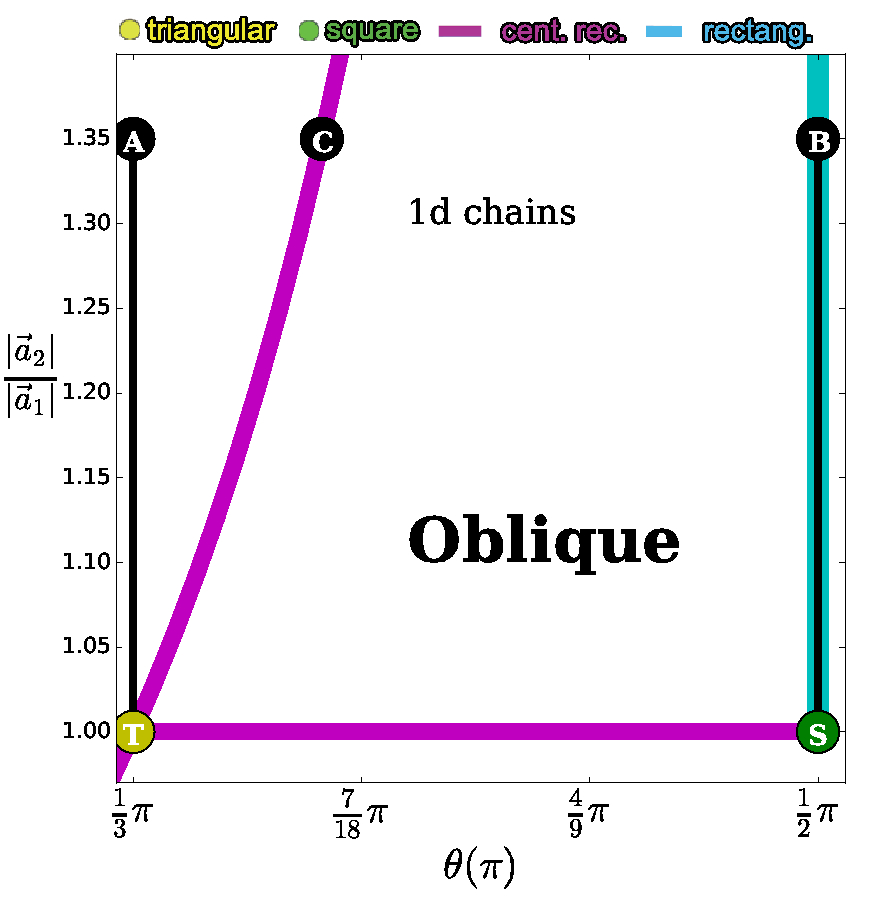}
\caption{(Color online) The space of all possible 2D Bravais lattices, parametrized as a function of the angle between the primitive vectors and the ratio of their length. Lattices of high symmetry are shown by lines and points.}\label{pointGroups}
\end{figure}

The most striking evolution is between the square lattice and the triangular lattice. One of the main diagonal gap in the square lattice energy spectrum shrinks, and after infinitely many gap closures and openings, the triangular lattice Hofstadter butterfly emerges. These gap closings and re-openings through the evolution are necessary to connect gaps with different Chern numbers in the two limits. We find that even a small departure from the high symmetry points such as the Triangular lattice, causes drastic changes in the spectrum, resulting in sudden energy gap openings and closures. The square lattice's bipartite symmetry causes the reflection symmetry along $E=0$ line in the Hofstadter butterfly. This symmetry is swiftly broken during the evolution and completely fades away in the Triangular lattice limit. For the Triangular lattice, the area of the primitive cell and consequently the flux per plaquette is reduced by one half compared to the oblique lattice. Hence, the periodicity of energy spectrum as a function of the flux is doubled. 

The transition between the square lattice butterfly and the triangular lattice butterfly has previously been studied by Hatsugai and Kohmoto\cite{KohmotoTritoSqu} by considering a square lattice with NNN hopping. While the results obtained directly starting from the TB model are valuable, it is not straightforward to link the TB parameters with the lattice geometry and real space potential. In this work, we generate the TB parameters starting from a realistic optical lattice potential and relate lattice geometry directly to the energy spectrum. For the square to triangular lattice transition, our results are in agreement with Ref.\cite{KohmotoTritoSqu}.

The square to rectangular lattice transition demonstrates how the self similar spectrum emerges from a continuous band. For a rectangular lattice with extreme anisotropy, the hopping amplitude in one direction is suppressed and the system is equivalent to a collection of one-dimensional chains. In one dimension, an external magnetic field can be gauged away and has no effect on the spectrum. As the anisotropy of the rectangular lattice is reduced, the isolated one-dimensional chains are weakly connected. Treating this connection perturbatively, we observe how the self similar energy gaps are formed. A similar transition takes place between the triangular lattice and the centered rectangular lattices. We also analyze the oblique and the centered rectangular to the triangular lattice transitions.

Finally, we investigate the Chern numbers of the main gaps during the lattice evolution. We find that Chern numbers are always transferred by a multiple of $q$ when two bands connect and split. We argue that this is due to the $q$-fold degeneracy in momentum space causing $q$ Dirac cones to emerge simultaneously. We also investigate the behavior of Chern numbers on a closed path in the space of two dimensional Bravais lattices.

This article is organized as follows, section II introduces the real space potential and constructs a proper TB model description under magnetic field. The next section explains the diagonalization of the magnetic TB Hamiltonian and calculates of the corresponding energy spectra. In Section IV, all possible Bravais lattice transitions and the corresponding Hofstadter butterflies are discussed in detail. In Section V, the Bravais lattices are characterized by their Chern number and the parameter space of lattices are classified by set of Chern numbers for each magnetic flux per plaquette. In the conclusion, we give an overview of our results and discuss the experimental implications. 

\section{THE MODEL}
In this section, we propose a lattice potential which can be adjusted to form any Bravais lattice in two dimensions. We reduce this potential to a TB model in the deep potential limit. The effect of a magnetic field is introduced by the Peierls substitution\cite{Peierls} in the next section. 

A 2D Bravais lattice is described by its primitive vectors\cite{Ashcroft},
\begin{eqnarray}\label{primitVecsDmnslss}
  \vec{a}_1 &=& \lambda_1 \hat{x}, \nonumber\\
  \vec{a}_2 &=& \lambda_2 \left( \cos{\theta} \hat{x} + \sin{\theta} \hat{y} \right),
\end{eqnarray}
where $\theta$ is the angle between the primitive lattice vectors, the lattice constants are $\lambda_1$ and $\lambda_2$. Without loss of generality, we choose $\vec{a}_1$ to be along the $\hat{x}$ direction. 

The Bravais lattice points are at positions,
\begin{eqnarray}\label{BravaisVecs}
\vec{R}_{n_1,n_2} &=& n_1 \vec{a}_1 + n_2 \vec{a}_2, \quad n_1,n_2 \in \mathbb{Z},\nonumber\\
  				&=& \big[(\lambda_1 n_1 + \lambda_2 n_2 \cos{\theta}) \hat{x} + (\lambda_2 n_2 \sin{\theta})\hat{y}\big].
\end{eqnarray} 
A real space potential $V(x,y)$ which is capable of generating all possible two-dimensional Bravais lattices with minimum number of Fourier components is,
\begin{eqnarray}\label{potentialEqn}
V(x,y)&=& - V_X cos (\vec{k}_1\cdot\vec{x}) - V_Y cos (\vec{k}_2\cdot\vec{x}) \nonumber  \\
     &-&  2 \cos{\theta} \sqrt{V_X V_Y} \cos (\left( \vec{k}_1 - \vec{k}_2 \right)\cdot\vec{x}).
\end{eqnarray}
Sinusoidal potentials are routinely produced in cold atom experiments by a retro reflected lasers. Two dimensional optical lattices are generated by using at least two beams. In principle, the above potential can be produced by at most three laser beams, but if $|k_1| = |k_2|$, the last term in the potential results from the interference of the first two. The laser wave vectors are $\vec{k}_1 = \frac{2 \pi}{\lambda_1 \sin{\theta}}(0,1) $, $\vec{k}_2 = \frac{2 \pi}{\lambda_2 \sin{\theta}}(- \sin{\theta},\cos{\theta})$. 

Up to a scale transformation, all two dimensional Bravais lattices can be formed as $|\lambda_1/\lambda_2|$ and $\theta$ are varied. The Schr\"{o}dinger equation for the above potential is,
\begin{equation*}
\Big[ -\frac{\hbar^2}{2m} \big(\partial^2_x + \partial^2_y \big) + V(x,y) \Big] \psi(x,y) = E \psi(x,y).
\end{equation*}
We do not introduce the magnetic field into the continuum equation as the translation symmetry group under a magnetic field is more complicated than the usual crystal symmetries\cite{ZakMagneticTranslationGroup}. Instead, we first project the continuum problem onto the lowest band in k-space forming a TB description. Such a description is not only accurate in the deep lattice limit but also easily adapted to include the external magnetic field by Peierls substitution. 

The TB parameters can be obtained by two methods. The first method is to calculate them by fitting the energy band obtained from TB model with the lowest energy band of the numerical solution of the continuum problem. In the second method, WFs for the lowest energy band are constructed and used to project the continuum problem to the TB model. If the WFs are obtained from the numerical solution of the continuum problem by direct Fourier transformation, these methods are equivalent. However, we use an alternative definition of WFs\cite{Kivelson} which facilitates direct construction of these functions from a finite system. We use both methods for all our lattices and find that the methods are in good agreement in the deep lattice limit.

The usual definition of WFs has a phase ambiguity. This gauge choice is usually resolved by requiring maximum localization which increases the computational burden\cite{WannierOriginal}. An alternative definition of WFs was given by Kivelson\cite{Kivelson} based on projections to the single band Hilbert space. The projected position operators for the $n^{th}$ band are defined as,
\begin{eqnarray*}
  \hat{x}_n &=& \hat{P}_n \hat{x} \hat{P}_n, \nonumber \\
  \hat{y}_n &=& \hat{P}_n \hat{y} \hat{P}_n, 
\end{eqnarray*}
where 
\begin{equation}
\hat{P}_n = \sum_{k}^{BZ} | n,k\rangle \langle n,k |,
\end{equation}
The eigenstates of $\hat{x}_n$, $\hat{y}_n$ are the WFs and the eigenvalues are the corresponding Wannier centers,
\begin{equation*}
    \hat{x}_n | W_n (\vec{r}-\vec{R}) \rangle  =  \vec{R} | W_n (\vec{r}-\vec{R}) \rangle.
\end{equation*}
This definition reduces to the usual Fourier transform WF definition, but is equally applicable to finite or disordered systems. Thus, we use this definition on a finite system to generate WF and calculate the TB parameters. 

We use a finite system with four unit cells along each primitive vector. The projection operator for the lowest band is formed by the first sixteen nearly degenerate eigenstates. Instead of separately diagonalizing $\hat{x}_n$, $\hat{y}_n$ one after another, we diagonalize a linear combination, say $\hat{O}_n = \hat{x}_n + \alpha \hat{y}_n$ with arbitrary $\alpha$ and obtain the corresponding eigenstates. Surprisingly, even a 4-by-4 finite lattice is large enough to capture the infinite lattice hopping parameters within one percent error.

\begin{figure}
\includegraphics[width=0.42\textwidth]{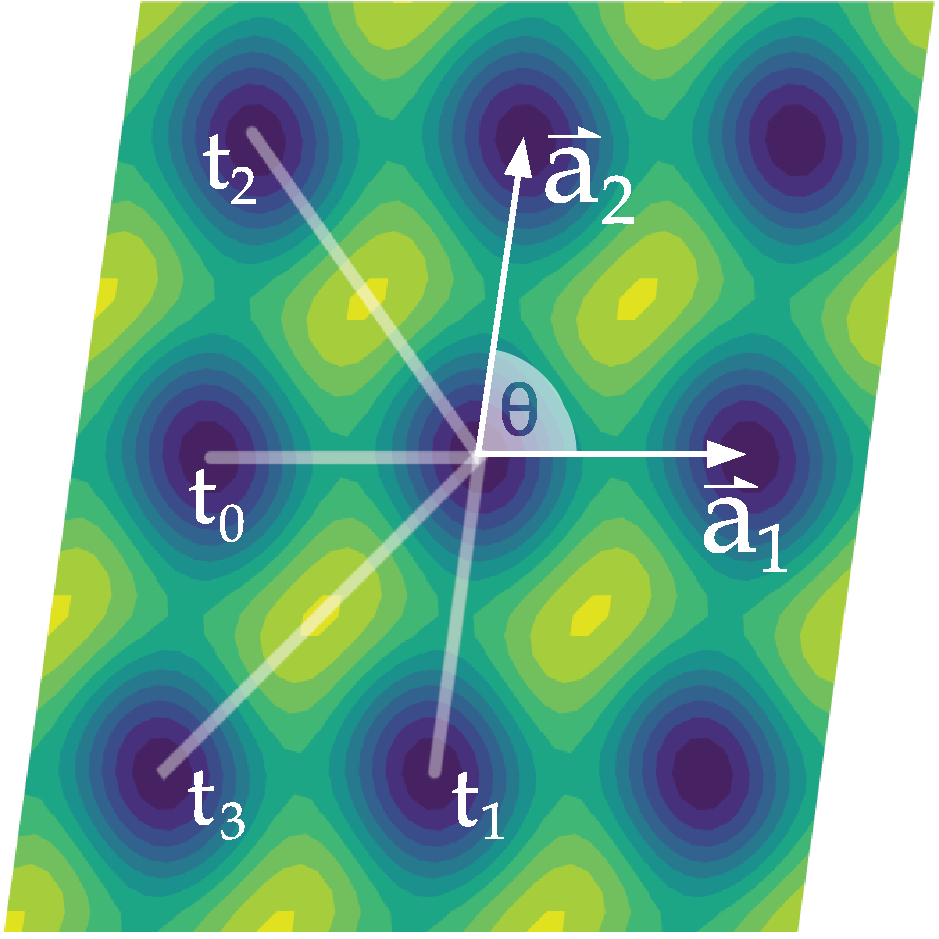}
\caption{(Color online) The TB model with eight neighbors represented on the contour plot of the potential. The bold arrows are the primitive translation vectors. The TB model hopping parameters are denoted with four different symbols, $t_0,t_1,t_2$ and $t_3$.}\label{tightBindModel}
\end{figure}

During the evolution of the lattice the number of nearest neighbors (NN) and the distance between NNs and next nearest neighbors (NNN) changes. In order to capture the physics of the transition, we calculate the TB parameters for eight neighbors shown in Fig.\ref{tightBindModel}. Due to inversion symmetry, only four distinct parameters are required.

The TB Hamiltonian for this system is,
\begin{eqnarray}\label{HamiltonianNoB}
{\cal \hat{H}} = - \sum_{m_1,m_2} \Big[&& t_0 | m_1+1  ,m_2  \rangle \langle m_1, m_2| \nonumber\\
&+& t_1 | m_1  ,m_2+1  \rangle \langle m_1, m_2| \nonumber\\
&+& t_2 | m_1-1,m_2+1 \rangle \langle m_1, m_2| \nonumber\\
&+& t_3 | m_1+1,m_2+1\rangle \langle m_1, m_2| \nonumber\\
&+& h.c.  \Big].
\end{eqnarray}
The Schr\"{o}dinger equation ${\cal \hat{H}} | \Psi \rangle = E |\Psi \rangle$ can be expressed in the localized basis $|\Psi \rangle= \sum_{n_1,n_2} \psi_{n_1,n_2} |n_1,n_2\rangle$. Then the Hamiltonian is diagonalized by a discrete Fourier transform, yielding the eigenvalues,
\begin{eqnarray}\label{EnergyDispersionNoB}
E(k_1,k_2)  &=& -2 t_0 cos(k_1) - 2 t_1 cos(k_2) \nonumber \\
            &-& 2 t_2 cos(k_1-k_2) - 2 t_3 cos (k_1 + k_2).
\end{eqnarray}
The dimensionless wavenumbers are, $k_1 = \vec{k} \cdot \vec{a_1}$ and $k_2 = \vec{k} \cdot \vec{a_2}$. This function reduces to TB bands for the square and the triangular lattice with corresponding NN and NNN hopping parameters.

We introduce the magnetic field using the Peierls substitution
\begin{equation*}
t_{\mathbf{m},\mathbf{n}} |\vec{R}_{\mathbf{n}} \rangle \langle \vec{R}_{\mathbf{m}} | \rightarrow e^{i \Theta_{\mathbf{m},\mathbf{n}}} t_{\mathbf{m},\mathbf{n}} |\vec{R}_{\mathbf{n}} \rangle \langle \vec{R}_{\mathbf{m}} |,
\end{equation*}
where $\mathbf{m} = (m_1,m_2)$ and $\mathbf{n} = (n_1,n_2)$. We choose the magnetic vector potential in the Landau gauge along $\vec{a}_1$ direction, $\vec{A} = B y \hat{x}$. The hopping phases are calculated as,
\begin{eqnarray*}
\Theta_{\mathbf{m},\mathbf{n}} &=& -\frac{e}{\hbar} \int_{\vec{R}_{\mathbf{m}}}^{\vec{R}_{\mathbf{n}}} \vec{A} \cdot \vec{d\ell} \\
            &=& - 2 \pi \phi \big([\vec{R}_{\mathbf{n}}-\vec{R}_{\mathbf{m}}]\cdot \hat{x}  \big)\big(\frac{\vec{R}_{\mathbf{n}}+\vec{R}_{\mathbf{m}}}{2}\cdot \hat{y} \big),
\end{eqnarray*}
where $\phi = B\lambda_1 \lambda_2\sin{\theta} / \phi_0$ is magnetic flux per plaquette normalized to flux quantum $\phi_0 = h/e$. This method faithfully describes a uniform magnetic field as long as the zero-field description of the TB model holds\cite{OktelP}.

With the Peierls substitution, the TB Hamiltonian under a magnetic field becomes,
\begin{eqnarray}\label{BHam}
{\cal \hat{H}} &=& - \sum_{m_1,m_2} \Big[ \nonumber\\
& & t_0 e^{-i2\pi\phi m_2} | m_1+1  ,m_2  \rangle \langle m_1, m_2| \nonumber\\
&+& t_1 e^{-i2\pi \phi \cos{\theta} (m_2+1/2)} | m_1  ,m_2+1  \rangle \langle m_1, m_2| \nonumber\\
&+& t_2 e^{i2\pi\phi(1-\cos{\theta}) (m_2+1/2)} | m_1-1,m_2+1 \rangle \langle m_1, m_2| \nonumber\\
&+& t_3e^{-i2\pi\phi(1+\cos{\theta}) (m_2+1/2)} | m_1+1,m_2+1\rangle \langle m_1, m_2| \nonumber\\
&+& h.c.].
\end{eqnarray}
The magnetic field enters the TB Hamiltonian only through the parameter $\phi$ which is the magnetic flux through a primitive unit cell of the lattice. As the lattice geometry evolves, the area of the primitive unit cell will in general change. Therefore, it is possible to investigate the evolution under two different constraints. First, one can envision that a uniform magnetic field is acting on the system so that $\phi$ changes with lattice geometry. Second approach is to take $\phi$ to be constant during the evolution. We use the second approach as it is more relevant to cold atom experiments where the artificial magnetic flux is generated by modifying the hopping parameters and is not affected by the unit cell area. In the next section, we calculate the energy spectrum under a constant $\phi$ as the geometry evolves.

\section{CALCULATION OF THE ENERGY SPECTRUM}
The magnetic Hamiltonian in Eq.\ref{BHam} acting on the state $|\Psi \rangle= \sum_{n_1,n_2} \psi_{n_1,n_2} |n_1,n_2\rangle$, yields the following difference equation,
\begin{eqnarray*}
E\psi_{m_1,m_2}&& = \nonumber \\
&-& t_0 \big(e^{-i2 \pi \phi m_2} \psi_{m_1-1,m_2} + e^{i2 \pi \phi m_2} \psi_{m_1+1,m_2}\big)    \nonumber \\
&-& t_1 \big(e^{-i2 \pi \phi \cos{\theta}(m_2 - 1/2)} \psi_{m_1,m_2-1} \nonumber \\
&& \quad + e^{i2 \pi \phi \cos{\theta}(m_2 + 1/2)} \psi_{m_1,m_2+1} \big) \nonumber \\
&-& t_2 \big(e^{i2 \pi \phi (1-\cos{\theta})(m_2 - 1/2)} \psi_{m_1+1,m_2-1} \nonumber\\
&& \quad  +e^{-i2 \pi \phi (1-\cos{\theta})(m_2 + 1/2)} \psi_{m_1-1,m_2+1} \big) \nonumber\\
&-& t_3 \big(e^{-i2 \pi \phi (1+\cos{\theta})(m_2 - 1/2)} \psi_{m_1-1,m_2-1} \nonumber \\
&& \quad  +e^{i2 \pi \phi (1+\cos{\theta})(m_2 + 1/2)} \psi_{m_1+1,m_2+1} \big).
\end{eqnarray*}
The Landau gauge, $\vec{A}$ is parallel to $\vec{a}_1$ and the Hamiltonian consequently preserves zero-field discrete translational symmetry in this direction. We choose a superposition of plane waves as the the mutual eigenstates of the Hamiltonian and the discrete translation operator along $\vec{a}_1$,
\begin{equation}
\psi_{m_1,m_2} (k_1,k_2)= e^{i k_1 m_1} g_{m_2}(k_1,k_2).
\end{equation}
With this choice, we obtain a one-dimensional difference equation for $g_{m_2}$. This equation is periodic only for rational values of $\phi$. Assuming $\phi = \frac{p}{q}$, where $p$ and $q$ are mutually prime integers, the one-dimensional equation becomes,
\begin{eqnarray*}
Eg_{m_2} =&-& t_0 \big(e^{-i2 \pi \frac{p}{q} m_2} e^{-ik_1} g_{m_2} \nonumber\\
&& \quad  + e^{i2 \pi \frac{p}{q} m_2} e^{ik_1} g_{m_2} \big) \nonumber\\
&-& t_1 \big( + e^{-i2 \pi \frac{p}{q} \cos{\theta}(m_2 - 1/2)}g_{m_2-1}  \nonumber\\
&& \quad + e^{i2 \pi \frac{p}{q} \cos{\theta}(m_2 + 1/2)} g_{m_2+1} \big) \nonumber\\
&-& t_2 \big(e^{i2 \pi \frac{p}{q} (1-\cos{\theta})(m_2 - 1/2)} e^{ik_1} g_{m_2-1} \nonumber\\
&& \quad + e^{-i2 \pi \frac{p}{q} (1-\cos{\theta})(m_2 + 1/2)} e^{-ik_1} g_{m_2+1} \big)\nonumber\\
&-& t_3 \big(e^{-i2 \pi \frac{p}{q} (1+\cos{\theta})(m_2 - 1/2)} e^{-ik_1} g_{m_2-1} \nonumber\\
&& \quad + e^{i2 \pi \frac{p}{q} (1+\cos{\theta})(m_2 + 1/2)} e^{ik_1} g_{m_2+1} \big).
\end{eqnarray*}
However, the periodicity of this difference equation by $q$ is not obvious. Following Rammal's approach\citep{Rammal}, we apply the following unitary transformation,
\begin{equation*}
g_{m_2} = e^{-i \pi \frac{p}{q} cos{\theta} m_{2}^2} f_{m_2}.
\end{equation*}
The periodicity of the resulting equation allows the use of Bloch's theorem. Hence, the diagonalization is reduced to a q-by-q matrix,
\begin{eqnarray}
E f_{m_2} = &-& t_0 \big(2 \cos{(2 \pi \frac{p}{q} m_2 + k_1)} f_{m_2} \big)\nonumber \\
&-& t_1 \big( f_{m_2-1}  + f_{m_2+1} \big) \nonumber \\
&-& t_2 \big(e^{i2 \pi \frac{p}{q}(m_2 - 1/2)} e^{ik_1} f_{m_2-1} \nonumber\\
&&\quad  \quad + e^{-i2 \pi \frac{p}{q} (m_2 + 1/2)} e^{-ik_1} f_{m_2+1} \big)\nonumber\\
&-& t_3 \big(e^{-i2 \pi \frac{p}{q} (m_2 - 1/2)} e^{-ik_1} f_{m_2-1} \nonumber\\
&& \quad \quad + e^{i2 \pi \frac{p}{q} (m_2 + 1/2)} e^{ik_1} f_{m_2+1} \big), \\
f_{m_2 + q} = && e^{i k_2 q} f_{m_2}.\nonumber
\end{eqnarray}
The calculation of the energies for all momenta $(k_1,k_2)$ within the magnetic BZ is computationally laborious. Instead of sampling the whole BZ, it is possible to identify the special $k$-points for which the energy values will be an extremum. Using symmetry to calculate these k-points makes it possible to obtain the band edges\citep{ChambersCri,WannierOriginal,Rammal} for each $\phi = p/q$ value by a few diagonalizations of a q-by-q matrix.

The reduced difference equation is solved numerically and we obtain the energy spectra for any lattice geometry. In the next section, we discuss these eigenvalue spectra and analyze the lattice transitions among the five Bravais lattices.

\section{TRANSITIONS BETWEEN BRAVAIS LATTICES}
In this section, we analyze the changes in the butterfly spectra calculated in the previous section as the lattice geometry between Bravais lattices of different symmetry adiabatically evolves. As discussed in the model section, we characterize all the Bravais lattices by two parameters, $\theta$ and $|\vec{a}_2|/|\vec{a}_1|$. In Fig.\ref{pointGroups}, we show the five lattices of distinct symmetry groups and the transition paths between them in the space of these two parameters. There are five Bravais lattices in two dimensions, the square, the triangular, the rectangular, the centered rectangular and the oblique lattices. The lattices with the higher symmetry occupy regions of smaller measure. In our representation, the whole area is covered by the oblique lattice which has the least amount of symmetry. The rectangular and the centered rectangular lattices correspond to one-dimensional curves while the most symmetric triangular and square lattices are confined to isolated points.
\begin{figure}
\hbox{\includegraphics[width=0.47\textwidth]{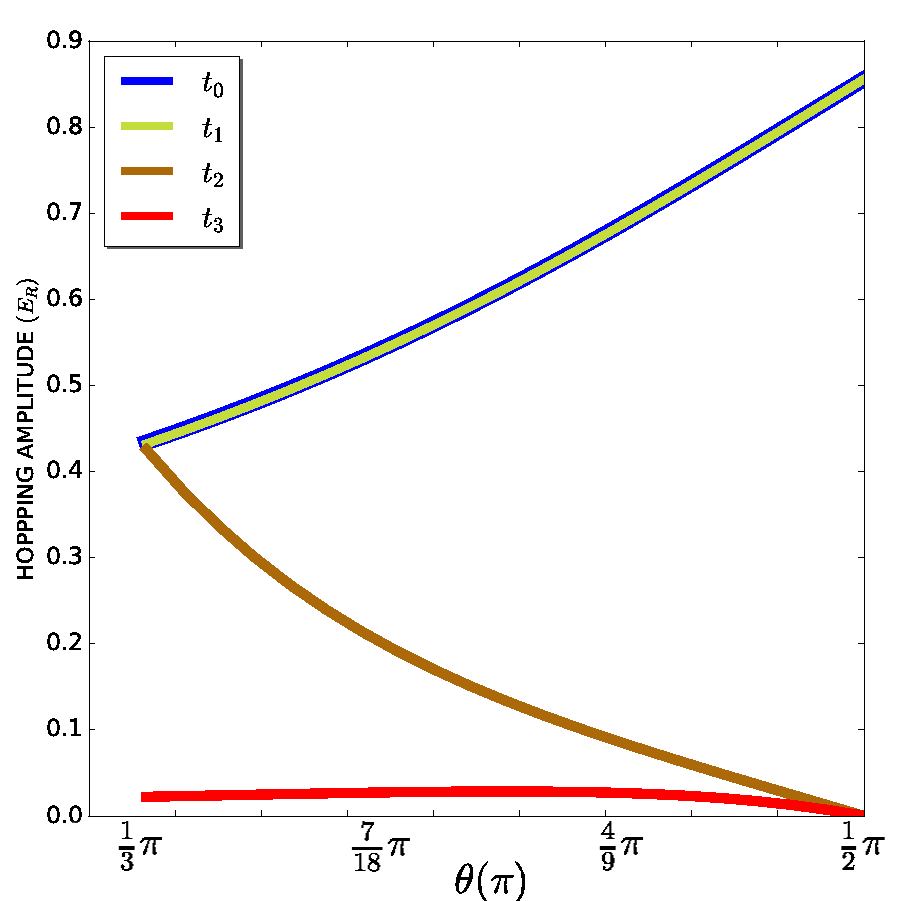}}
\caption{(Color online) The hopping amplitudes calculated along the transition between the square lattice and the triangular lattice as a function of $\theta$. The parameters are calculated for $V_X = V_Y = 20 (E_R)$. The NN hopping parameters, $t_0,t_1$ have the same magnitude and decrease until they are equal to $t_2$ in the triangular lattice limit. The NNN hopping, $t_3$ is an order of magnitude smaller.}\label{hopping}
\end{figure}
\begin{figure}
\hbox{\includegraphics[width=0.47\textwidth]{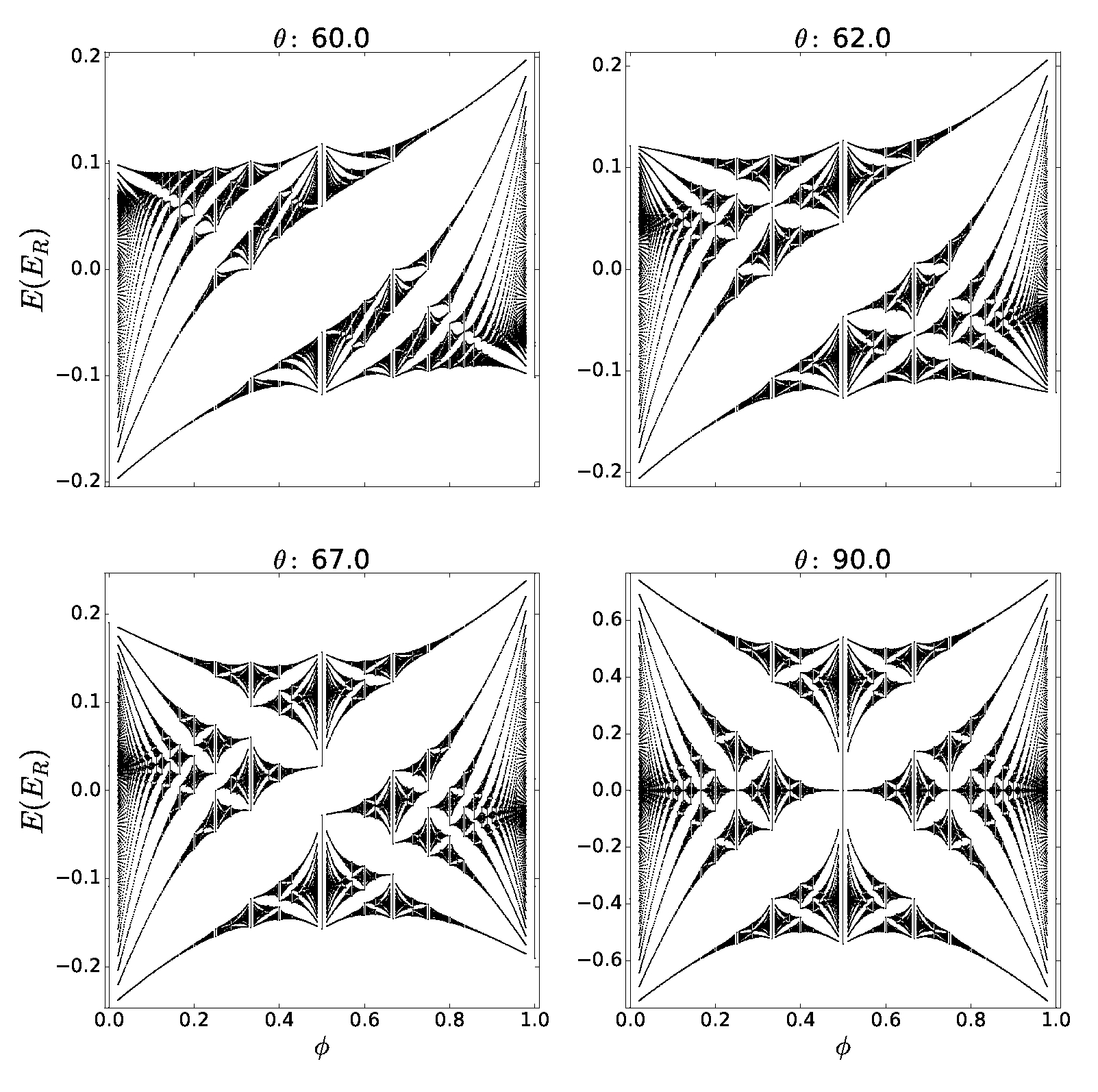}}
\caption{Evolution of the energy spectrum between the square lattice and the triangular lattice in four representative points. One of the major energy gaps disappears as the triangular lattice is approached. The closing is accompanied by infinitely many smaller gap closures. These results are in agreement with\citep{KohmotoTritoSqu}, but show that the transition in terms of the angle $\theta$ happens mainly near the triangular lattice limit.}\label{squTOtri}
\end{figure}

We start by considering the transition between the two most symmetric lattices, the square and the triangular lattices. In our parameter space Fig.\ref{squTOtri}, the path $\overline{BA}$ indicates this transition. The energy spectra at four representative points along that path are displayed in Fig.\ref{squTOtri}. 

During the square to the triangular lattice transition, the magnitude of the hopping elements are calculated over this path. For a real space potential with $V_X = V_Y = 50 (E_R)$ resembling the square lattice, the hopping parameters are $t_0 = t_1 = 0.1913$ and $t_2,t_3 \approxeq 0$. The four NN hopping amplitudes are equal while the four NNN hoppings almost vanish. This is an expected result in the deep lattice limit as the WFs are localized to within a unit cell and the NNN transitions are strongly suppressed. The interference term in the potential in Eq.\ref{potentialEqn} increases towards the triangular lattice. During the evolution, $t_0$ and $t_1$ decrease with the same magnitude until they are equal to $t_2 = 0.0345$ in the triangular lattice limit, see Fig.\ref{hopping}. It is natural that the NN elements are almost constant because only the angles between the local minima of two adjacent sites changes.  

The square lattice butterfly has bipartite symmetry when NNN hopping is neglected. Even when this symmetry is broken by weak NNN hopping, the spectrum approximately preserves the reflection symmetry along $E=0$ line. This symmetry is broken more strongly when the system moves towards the triangular lattice. The amount of the asymmetry is maximum at $\phi = 1/2$ where it is proportional to $|t_0-t_2|$. The asymmetry of the energies results in shrinking of the energy gaps at the upper right and the lower left main gaps of the butterfly. Meanwhile, the lower right and the upper left wings enlarge to form the triangular butterfly main skeleton. This lattice transition is non-trivial as infinitely many gaps close and reopen during the evolution. This is expected as the square lattice and the triangular lattice are adiabatically disconnected\citep{Avron2}. 

We observe that the energy spectrum is sensitively dependent on the parameters near the high symmetry points. Even small deviations of the lattice from these points result in large shifts of energy bands as well as gap closures. Hence, the triangular and the square lattice butterflies are representative of only small regions in the parameter space. This point is especially important for design of experimental lattices.
 
As discussed in the model section, we keep the flux per unit cell $\phi$ constant during the transition. The flux per plaquette drops to $\phi/2$ at the triangular lattice limit. This flux halving was first discussed by Claro and Wannier\cite{HofstadterWannier}. Consequently, the triangular lattice butterfly is periodic by $\phi = 2$ as seen in Fig.\ref{squTOtri}.
\begin{figure*}
\centerline{\includegraphics[width=1\textwidth]{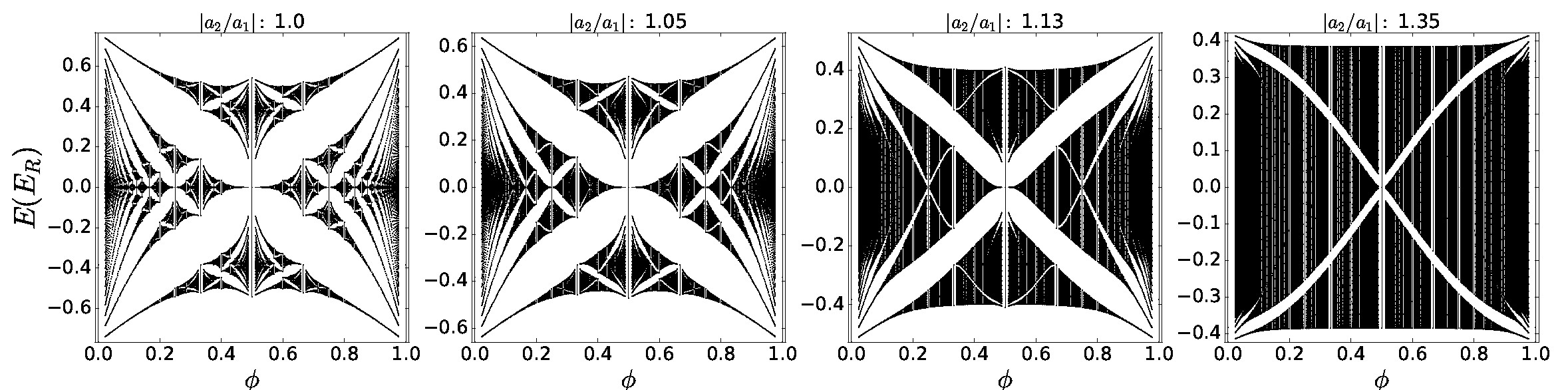}}
\centerline{\includegraphics[width=1\textwidth]{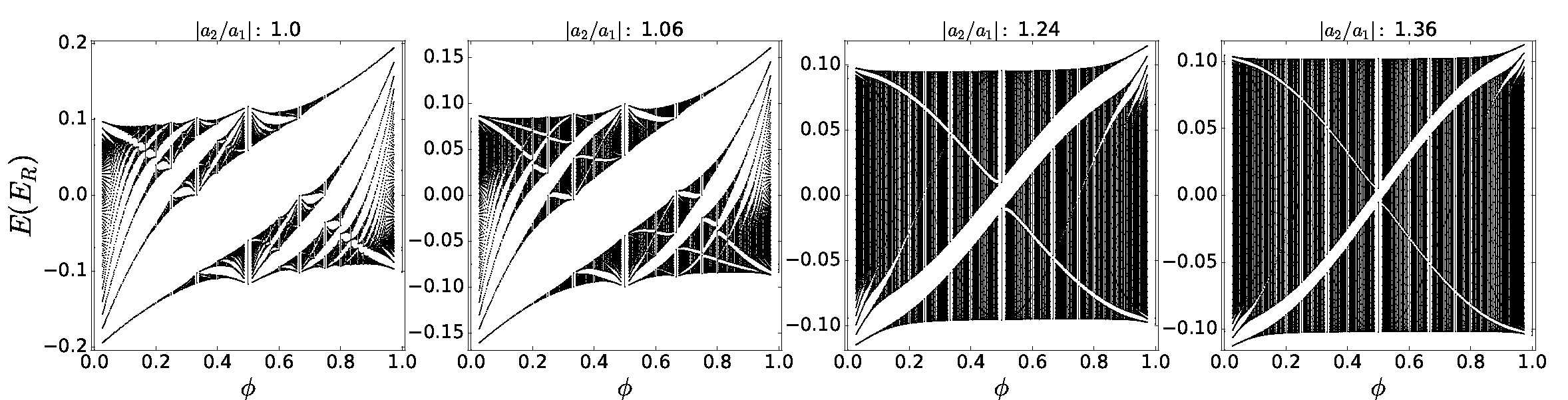}}
\caption{Evolution of the energy spectrum from the square lattice to the rectangular lattice, and from the triangular lattice to the centered rectangular lattice. The first transition is along $\theta = \pi/2$ line while the second transition is calculated along the centered rectangular lattice curve shown in Fig.\ref{pointGroups}. The greater the ratio of the primitive vectors, the closer the system is to one-dimensional chains (see Fig.\ref{rectGaps}), where the sub-band splitting is weak.}\label{squTOrec}
\end{figure*}

\begin{figure}
\includegraphics[width=0.47\textwidth]{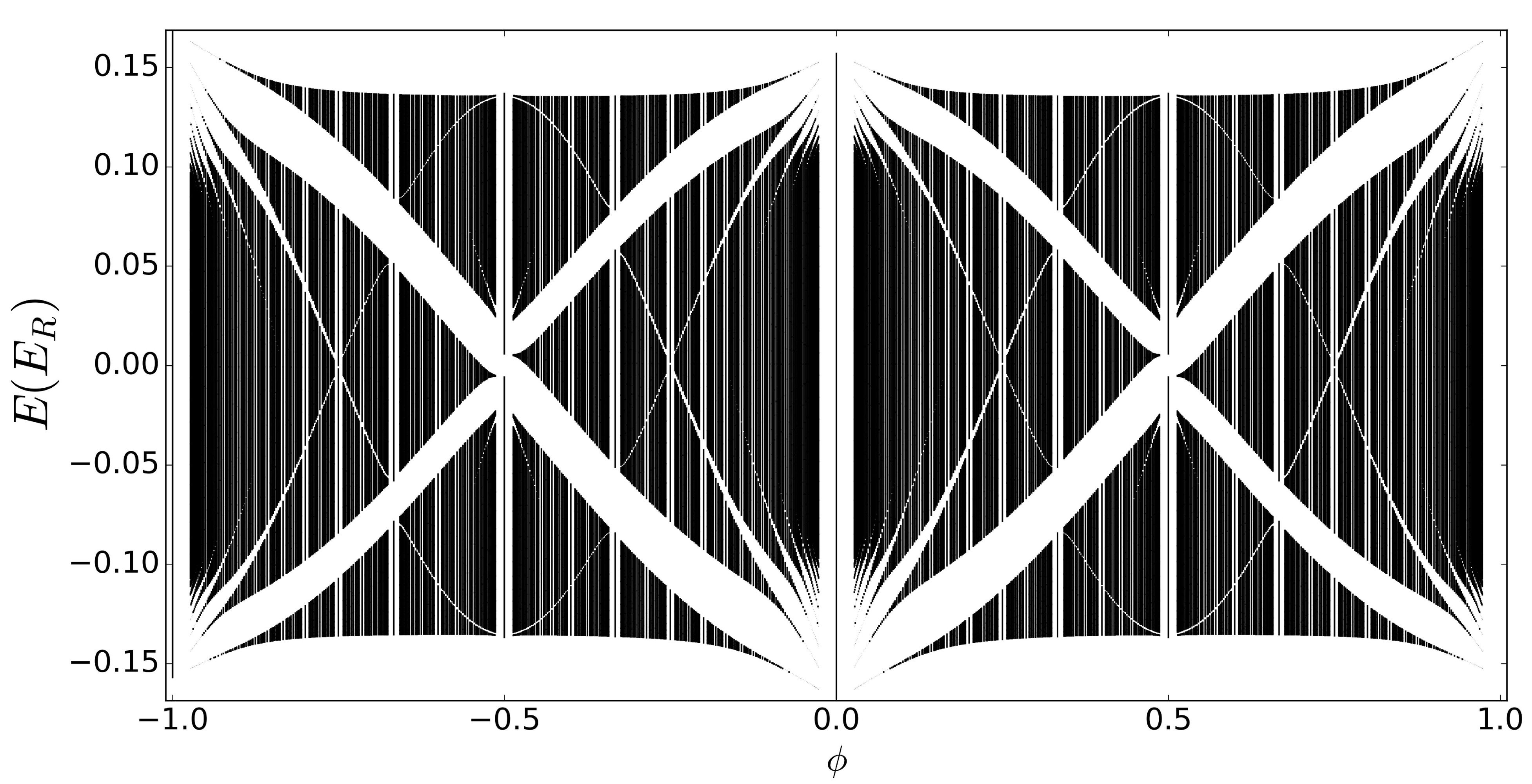}
\caption{A representative energy spectrum of the oblique lattice. We display the spectrum from $\phi =[-1,1]$ to emphasize the doubling of the $\phi$ periodicity and the symmetry under $\phi \to -\phi$. All 2D Bravais lattices have this symmetry which follows from inversion.}\label{obliquephiminusphi}
\end{figure}

The energy spectrum is symmetric under $\phi \to -\phi$ operation for all the Bravais lattices. This symmetry is easy to understand for all cases except the oblique geometry. If the lattice geometry does not distinguish between $\pm z$ axis, the resulting Hamiltonian is invariant under the reversal of the flux. For the oblique lattice, although $\pm z$ directions can be uniquely defined, the lattice still has inversion symmetry which coupled by a reflection in the lattice plane restores the $\phi \to -\phi$ symmetry as seen in Fig.\ref{obliquephiminusphi} and it is therefore valid for all Bravais lattices. The inversion symmetry can only be broken when there is an energy difference in on-site energies of NNs. 
\begin{figure}
\includegraphics[width=0.47\textwidth]{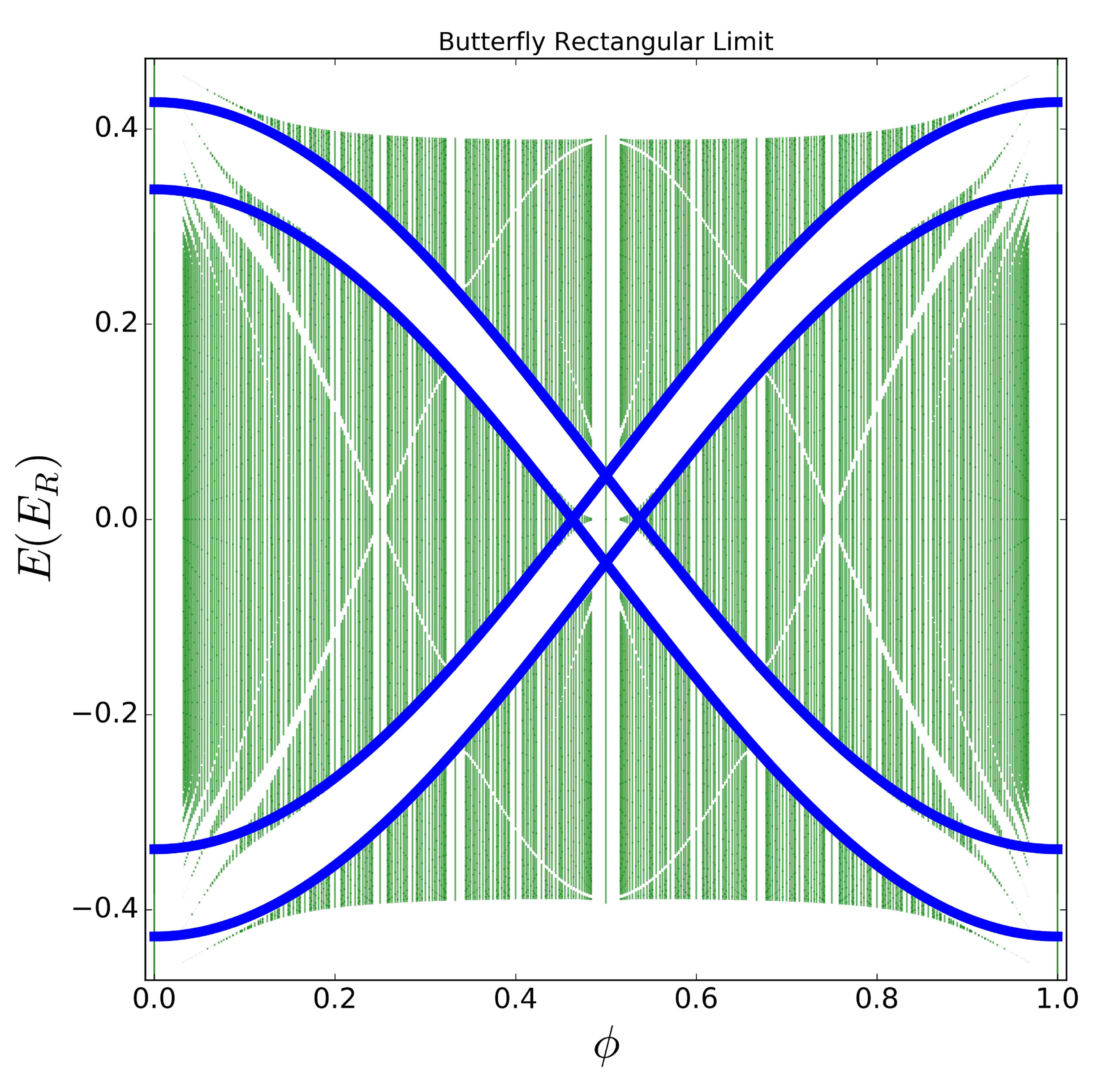}
\caption{(Color online) The energy gaps are calculated in the rectangular limit compared with perturbative treatment of 1D chains. In the first order, the spectrum has two robust energy gaps. The smaller gaps in the spectrum can be generated by higher orders in perturbative tunneling.}\label{rectGaps}
\end{figure}

The next transition we consider is between the square lattice and the rectangular lattice.  Although this transition happens as soon as four-fold symmetry is broken, it is instructive to study the evolution from the square limit to the extremely anisotropic case. If tunneling in one direction is much weaker than the other one, the system is a collection of one-dimensional chains. During this transition the square lattice butterfly energy gaps disappear and the spectrum becomes a one-dimensional TB band. It is not surprising that for a collection of 1D chains, the spectrum is independent of the magnetic field. For a 1D chain, the external magnetic field can be gauged out. Fig.\ref{rectGaps} shows the evolution of the energy spectrum till the rectangular lattice with extreme asymmetry (for $|\vec{a}_1/\vec{a}_2| = 2.5$ and the hopping amplitudes, $t_0 = 0.19(E_R)$, $t_1 = 0.04(E_R)$). The energy spectra are mainly divided into three bands and corresponding two main energy gaps are preserved for even for large $|\vec{a}_1/\vec{a}_2|$ values. 

The robustness of the major gaps can be understood by a perturbative approach. We consider the Hamiltonian of independent 1D chains as the unperturbed system and the hopping between the chains as the perturbation. The dispersion of the $m^{th}$ unperturbed chain is,
\begin{equation}
E_m(k_1)  = - 2 t_0 \cos{(2 \pi \phi m + k_1)}.
\end{equation} 
The energy bands of adjacent chains are degenerate at two k-points and separated by $2 \pi \frac{p}{q}$ for $\phi = \frac{p}{q}$. The tunneling amplitude between the adjacent chains, $t_1$ lifts the degeneracy at these k-points and the new energies become $\pm 2 t_0 \cos{(\pi \frac{p}{q} + k_1)} \pm t_1$. Therefore, $t_1$ is the measure of the energy splitting for the main gaps of the butterfly as in Fig.\ref{rectGaps}. The perturbative approach can also generate the minor gaps and the Hofstadter butterfly emerges when higher orders are included\cite{FracHallOneDimChains}.

We conclude this section with the discussion of transitions from the triangular lattice to the centered rectangular and to the highly anisotropic oblique lattice. These two transitions are represented in Fig.\ref{pointGroups} by the paths $\overline{AG}$, $\overline{AE}$. In both cases, the evolution of the energy spectrum is similar to the transition from the square to the rectangular lattice. Starting from the triangular lattice butterfly, the smaller gaps close first and then the larger gaps shrink to form an energy spectrum of weakly coupled one-dimensional chains. Due to the lack of bipartite symmetry, only one of the major gaps is robust unlike the square to rectangular transition. The asymmetry between the two diagonal gaps can be used as a measure of the closeness of the system to bipartite symmetry. 
\begin{figure}
\hbox{\includegraphics[width=0.47\textwidth]{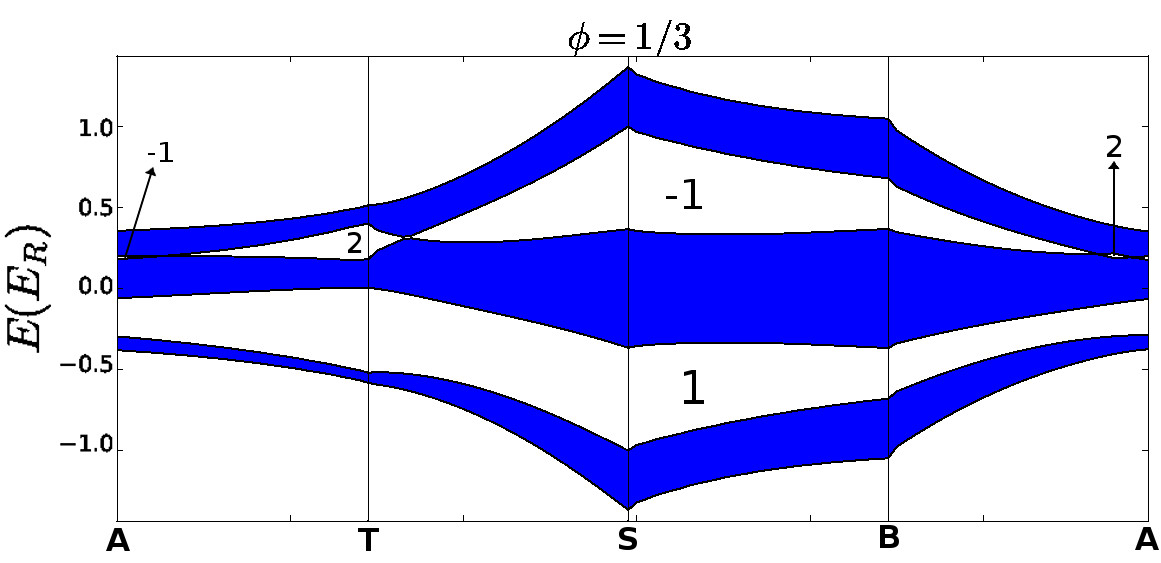}}
\hbox{\includegraphics[width=0.47\textwidth]{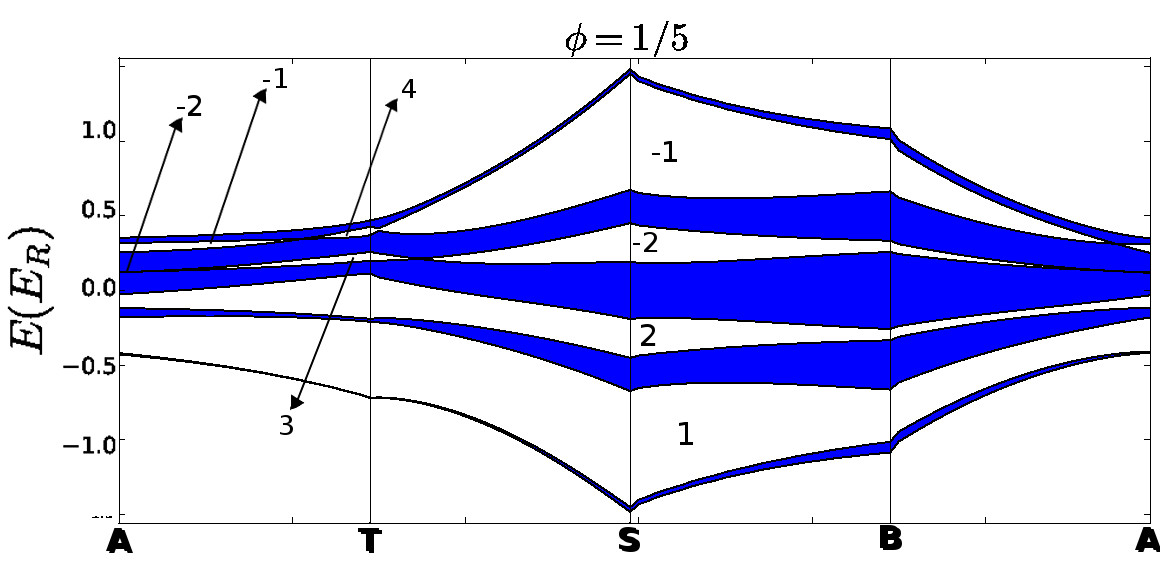}}
\caption{(Color online) The evolution of the energy bands for $\phi=\frac{1}{3}$ and $\phi=\frac{1}{5}$ on the path $\overline{ATSBA}$ shown in Fig.\ref{pointGroups}. Gaps are labeled with their Chern numbers. For $\phi=1/3$, there are two topologically distinct regions, characterized by $1,-1$ and $1,2$. For $\phi=1/5$, there are four topologically distinct regions.}\label{phiEvolution}
\end{figure}

\begin{figure*}
\centerline{\includegraphics[width=1\textwidth]{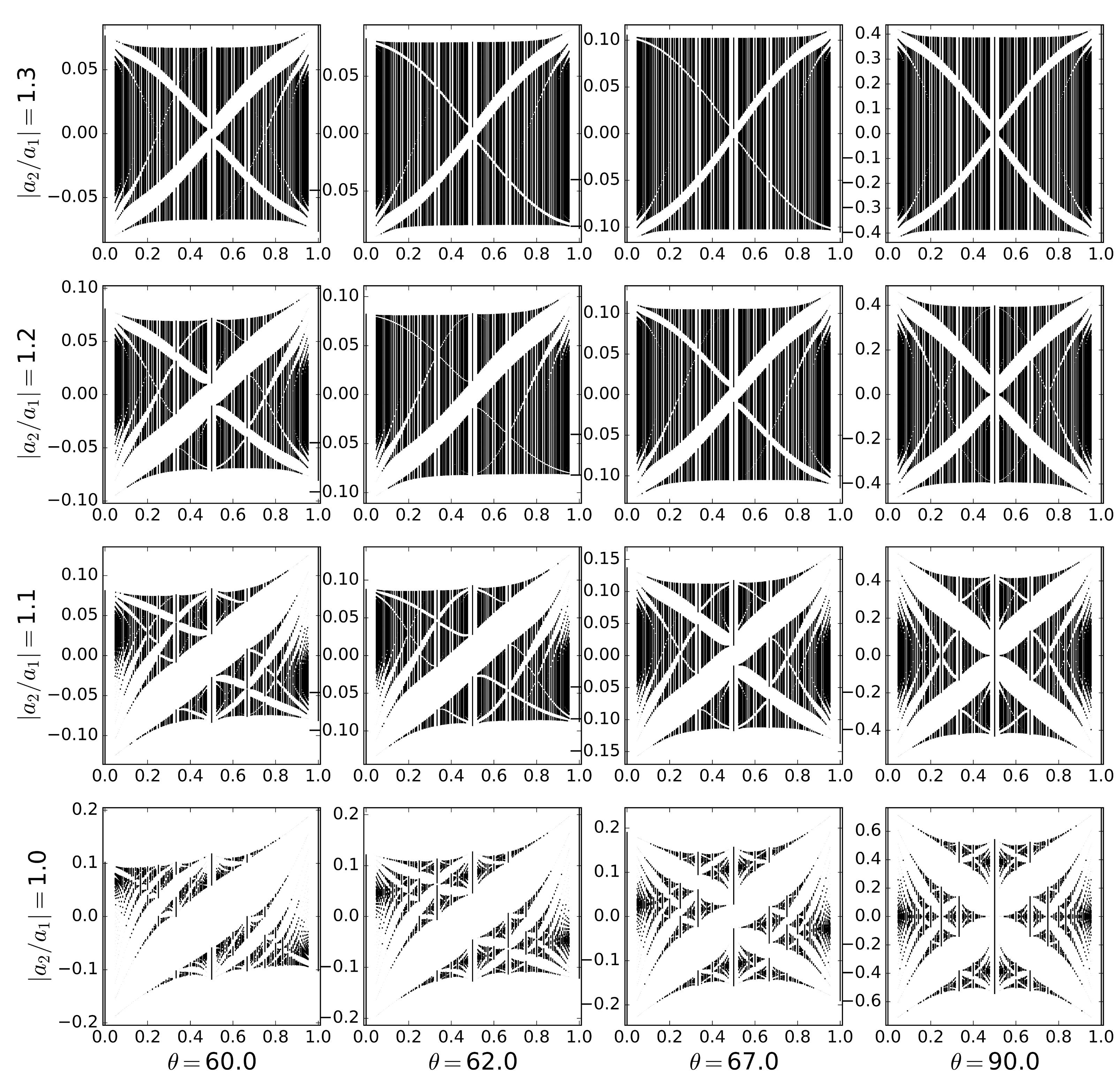}}
\caption{An overal picture of the evolution of the Hofstadter butterfly in the phase space of two dimensional Bravais lattices.}\label{regional}
\end{figure*}

All possible energy spectra for the Bravais lattice parameter space is depicted in Fig.\ref{regional}.

In our TB model, the smallest plaquette area that can be enclosed by hopping is half of the unit cell. Therefore, all the energy spectra we calculate are periodic by $\phi = 2$. However, in the cases where NNN hoppings ($t_2,t_3$) are negligible the smallest enclosed area is the unit cell and the spectra are periodic with $\phi =1$. In general, the periodicity of the spectrum with $\phi$ can be used to determine how long the range of significant hopping is. In the centered rectangular and the oblique lattice limits, the energy spectra preserve the periodicity of $\phi = 2$ magnetic flux. The smallest cells in each case are an half of the unit cells and therefore the enclosed flux are halved as well.

\section{Topological Characterization of the Bravais Lattice Phase Space}
The energy bands of lattices under magnetic field are characterized by a topological invariant, the first Chern number\citep{TKNN}. The contribution of a band to the Hall conductivity for a non-interacting fermionic system is given by the Chern number, $\sigma_{xy} = \mathcal{C} e^2/h$. Alternatively, Chern numbers can be associated with gaps by summing Chern numbers of all the bands below a certain gap. In this section, we use the Chern numbers to characterize the energy spectrum as the geometry of the lattice changes.

For a lattice with flux $\phi = p/q$, the energy spectrum will have $q$ bands and $q-1$ gaps. While the number of the gaps are determined only by the flux, the Chern numbers associated with these gaps change with lattice geometry. For example, for $\phi =1/3$, the square lattice gaps have Chern numbers $1,-1$, while the triangular lattice gaps have $1,2$. These Chern numbers would not be affected by small changes in the geometry as they are topologically protected. Chern numbers only change if a gap in the spectrum closes. Thus, we can classify all lattices into equivalence classes by their Chern number sequence. Hence, our parametrization of the lattice space in Fig.\ref{pointGroups} can be separated into regions of topologically equivalent phases. The phase boundaries, then, show the parameters for which at least one gap closes.

Chern numbers can be calculated by k-space integration of Berry curvature over the BZ. However, we use an indirect but computationally simpler approach\cite{OnurHuiOktel}. The conductivity of the system is a thermodynamic variable and can be calculated as a variation of the number of levels below the Fermi level with respect to the changes in the magnetic field in two dimensions\citep{Streda}. In lattice systems, the Chern number for a gap is,
\begin{equation}
\mathcal{C} = \frac{\partial n}{\partial \phi}
\end{equation}
where $n$ is the density per unit cell for states below the Fermi level, $\phi$ is the magnetic flux per plaquette. Chern number is calculated through the difference of the energy eigenvalues below the Fermi energy for two close magnetic flux values.

\begin{figure}
\includegraphics[width=0.47\textwidth]{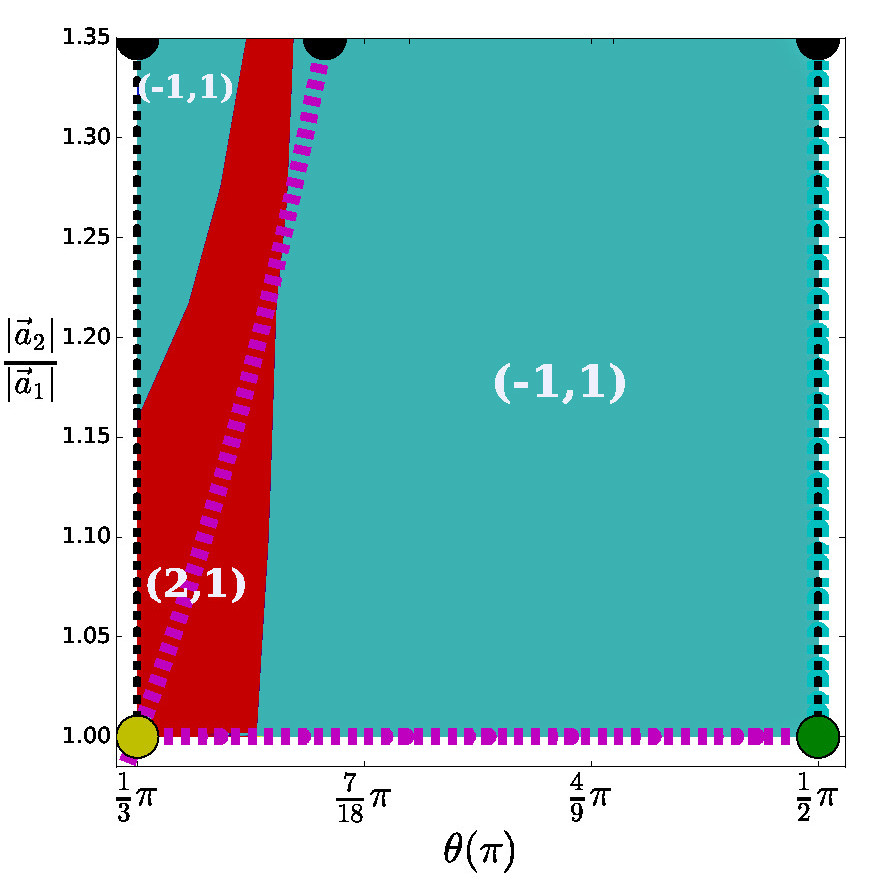}
\caption{(Color online) The topological regions of the two dimensional Bravais lattice phase space for $\phi = 1/3$. All lattices can be classified into two regions; those with Chern numbers $-1,1$ including the square lattice and those with $2,1$ including the triangular lattice. The triangular lattice region roughly tracks the centered rectangular lattice line.}\label{chernmap}
\end{figure}

We explored the lattice phase space for two simplest non-trivial fluxes, $\phi = 1/3,1/5$. First, we take a closed rectangular path over the phase space, $ATSBA$ and calculate the evolution of the energy bands. The path is chosen such that it avoids the 1D chains limit where gaps are very small as in Fig.\ref{squTOrec}. These evolutions are given in Fig.\ref{phiEvolution}. For $\phi=1/3$, we find that all lattices are either equivalent to the square lattice with Chern numbers $-1,1$ or equivalent to the triangular lattice with $2,1$. Thus, we separate the parameter space into two regions as can be seen in Fig.\ref{chernmap}. We find that the region equivalent to the triangular lattice roughly tracks the centered rectangular lattice curve. For $\phi=1/5$, we find four different regions which are sampled by the rectangular cut in Fig.\ref{phiEvolution}. As $q$ is increased, we get smaller gaps and correspondingly the phase diagram splits into smaller regions.

An important observation is that at the phase boundaries, the Chern numbers for energy gaps change exactly by $q$ for magnetic flux $\phi = p/q$. It is possible to understand this surprising observation by a symmetry argument. Because of the magnetic translation symmetry, the energy spectrum is $q$-fold degenerate inside the magnetic BZ. When two bands touch they must be degenerate at least $q$ points. A Chern number of exactly $1$ is exchanged when a single Dirac cone closes and re-opens.  If each one of the degenerate points is a Dirac cone, total Chern number of $q$ is exchanged between the bands. If degenerate points are of higher order, integer multiples of $q$ can be exchanged. However, we have not numerically observed such an exchange.

\section{CONCLUSION}
The cold atom experiments in optical lattices pioneered a vast number of interesting phenomena, which are not possible to realize in solid state systems. Especially, the enhanced control over the lattice geometry and the artificial gauge fields can be utilized to directly realize the Hubbard-Hofstadter model\cite{KetterleHarper,BlochHofstadter}.

In this respect, we investigate the inherent relation between the point group symmetries in two dimensions and the energy spectrum as a function of magnetic flux per plaquette. We focus on the transitions between lattices of different symmetry by using an optical lattice potential, which can realize all the two dimensional Bravais lattices. We describe this potential through a TB model and calculate its parameters ab-initio. The effect of the magnetic field is introduced by the Peierls substitution. Then, the energy spectra for each symmetry group are calculated as well as the evolution of the energy spectrum during transition between the symmetries.

We find that the lattice deformations around high symmetry points yield dramatic changes in the energy spectra. The evolution between the square lattice and the triangular lattice is mainly influenced by broken bipartite symmetry. We find that the energy spectrum changes dramatically in the vicinity of $\theta = \pi/3$,  the triangular lattice limit. This rapid change should be experimentally observable even for the simplest non-trivial flux value, $\phi = 1/3$. A few degrees of deviation from $\theta = \pi/3$ is found to make a jump in Hall conductivity if the system is filled up to the first gap. 

During the evolution, bands touch and reopen to transfer Chern numbers between the gaps. We find that the Chern number of bands only change with integer multiples of $q$. We explain this observation by invoking the $q$-fold degeneracy within a magnetic BZ. This result is particularly important for solid state experiments where Hall conductivity is directly measured through transport.

Finally, we regard the space of all possible Bravais lattices as a phase diagram where each phase is identified by $q-1$ Chern numbers of the gaps. For $\phi = 1/3$, there are two distinct regions belonging to the square lattice and the triangular lattice. In addition, the Chern number map for the phase space indicates that the region corresponding to the triangular lattice is found to roughly follow the centered rectangular lattice curve.  For $\phi = 1/5$, the space is divided into four phases and larger values of $q$ result in smaller topological regions.

\begin{acknowledgments}
F.Y. is supported by T\"{u}rkiye Bilimsel ve Teknolojik Ara\d{s}t{\i}rma Kurumu (T\"{U}B\.{I}TAK) Scholarship No. 2211.
\end{acknowledgments}

\bibliography{arxiv_24_march_v1}% Produces the bibliography via BibTeX.
\end{document}